\documentclass[aps,pre,twocolumn,superscriptaddress,showpacs,floatfix,nofootinbib]{revtex4-2} 
\usepackage{dcolumn}
\usepackage{amsmath}
\usepackage{amssymb}
\usepackage{amsthm}
\usepackage{graphicx}
\usepackage{bm}
\usepackage[T1]{fontenc}
\usepackage{color}
\usepackage{url}
\usepackage[bf]{subfigure}
\usepackage{rotating}
\usepackage{multirow}

\usepackage{scalefnt}
\usepackage{times}
\usepackage{mathtools}
\usepackage{bbm}

\usepackage{float}
\usepackage{enumerate}
\usepackage[makeroom]{cancel}

\usepackage{tikz}
\usetikzlibrary{topaths,calc}

\usepackage[hidelinks]{hyperref}

\newcommand{\E}[1]{\left\langle #1 \right\rangle}

\begin{document}
\title{Contagion dynamics on higher-order networks}

\author{Guilherme Ferraz de Arruda}
\affiliation{CENTAI Institute, Turin, Italy}

\author{Alberto Aleta}
\affiliation{Institute for Biocomputation and Physics of Complex Systems (BIFI), University of Zaragoza, Zaragoza, Spain}
\affiliation{Department of Theoretical Physics, Faculty of Sciences, University of Zaragoza, Zaragoza, Spain}

\author{Yamir Moreno}
\affiliation{Institute for Biocomputation and Physics of Complex Systems (BIFI), University of Zaragoza, Zaragoza, Spain}
\affiliation{Department of Theoretical Physics, Faculty of Sciences, University of Zaragoza, Zaragoza, Spain}
\affiliation{CENTAI Institute, Turin, Italy}


\begin{abstract}
   Understanding the dissemination of diseases, information, and behavior stands as a paramount research challenge in contemporary network and complex systems science. The COVID-19 pandemic and the proliferation of misinformation are relevant examples of the importance of these dynamic processes, which have recently gained more attention due to the potential of higher-order networks to unlock new avenues for their investigation. Despite being in its early stages, the examination of social contagion in higher-order networks has witnessed a surge of novel research and concepts, revealing different functional forms for the spreading dynamics and offering novel insights. This review presents a focused overview of this body of literature and proposes a unified formalism that covers most of these forms. The goal is to underscore the similarities and distinctions among various models, to motivate further research on the general and universal properties of such models. We also highlight that while the path for additional theoretical exploration appears clear, the empirical validation of these models through data or experiments remains scant, with an unsettled roadmap as of today. We therefore conclude with some perspectives aimed at providing possible research directions that could contribute to a better understanding of this class of dynamical processes, both from a theoretical and a data-oriented point of view.
\end{abstract}

\maketitle


\section{Introduction}

Contagion models cover a wide range of processes,  from the spread of diseases~\cite{barrat_2008, Pastor-Satorras2015, Arruda2018} to social contagion~\cite{Arruda2020} and rumor dissemination~\cite{daleykendall1964,makithompson1973}. These diverse processes have undergone extensive examination across various disciplines and under different perspectives. In the realms of physics and mathematics, contemporary approaches often incorporate heterogeneous interaction patterns~\cite{barrat_2008, Pastor-Satorras2015, Arruda2018}. However, these models traditionally assume pairwise interactions, encapsulated within graphs, limiting propagation to interactions between two individuals. This paradigm has recently been challenged, as new models have been proposed to account for group interactions using hypergraphs~\cite{battiston_networks_2020, bianconi_higher-order_2021, Torres2021, Battiston2021, Bick2023, Boccaletti2023}. In other words, the paradigm changes from the one-to-one formalism to the one-to-many or many-to-many interaction types. An illustrative example to gain some intuition is evident in modern messaging applications' group chats, which enable one-to-many interactions, in addition to direct messages between users.

In the literature, the introduction of these models typically justifies their inception by aiming to offer a more accurate depiction of specific social or epidemiological processes. However, most models remain predominantly theoretical, using these phenomena as inspirations for model definitions, yet lacking empirical validation. This stands in stark contrast to the advancement in the theoretical underpinnings of these processes, showcasing prolific growth. Various functional forms have been proposed to describe spreading in higher-order systems, accompanied by a diverse range of analytical techniques. These techniques span from classical approximations in graphs, such as heterogeneous mean-field approaches, to innovative methodologies like facet approximation~\cite{Jihye2023}. Thus, this review emphasizes both the generalities and specificities of these models, aiming to aid future research in generalizing contagion theory within higher-order networks, and with the aspiration that it will stimulate further empirical research.

To achieve this goal, in Sec.~\ref{sec:mot} we discuss studies and observations motivating the introduction of higher-order interactions in social and epidemic contagion models. Additionally, it outlines a set of open experimental questions that might be addressed with these approaches. Then, in Sec.~\ref{sec:Contagion}, we present a unified formalization covering the majority of models in the existing literature. Initially, we study this formulation in the limit where pairwise graphs are recovered, aiming to glean insights into such processes. Subsequently, the section is dedicated to an exploration of diverse approaches proposed for integrating higher-order interactions into these models, highlighting their similarities and distinctions. Concluding this review in Sec.~\ref{sec:Generalities}, we summarize major challenges and offer theoretical and data-oriented perspectives to pave the way for further research of higher-order contagion models.

\section{Phenomena motivating higher-order dynamics\label{sec:mot}}

\begin{figure*}[t!]
  \centering \includegraphics*[width=0.99\linewidth]{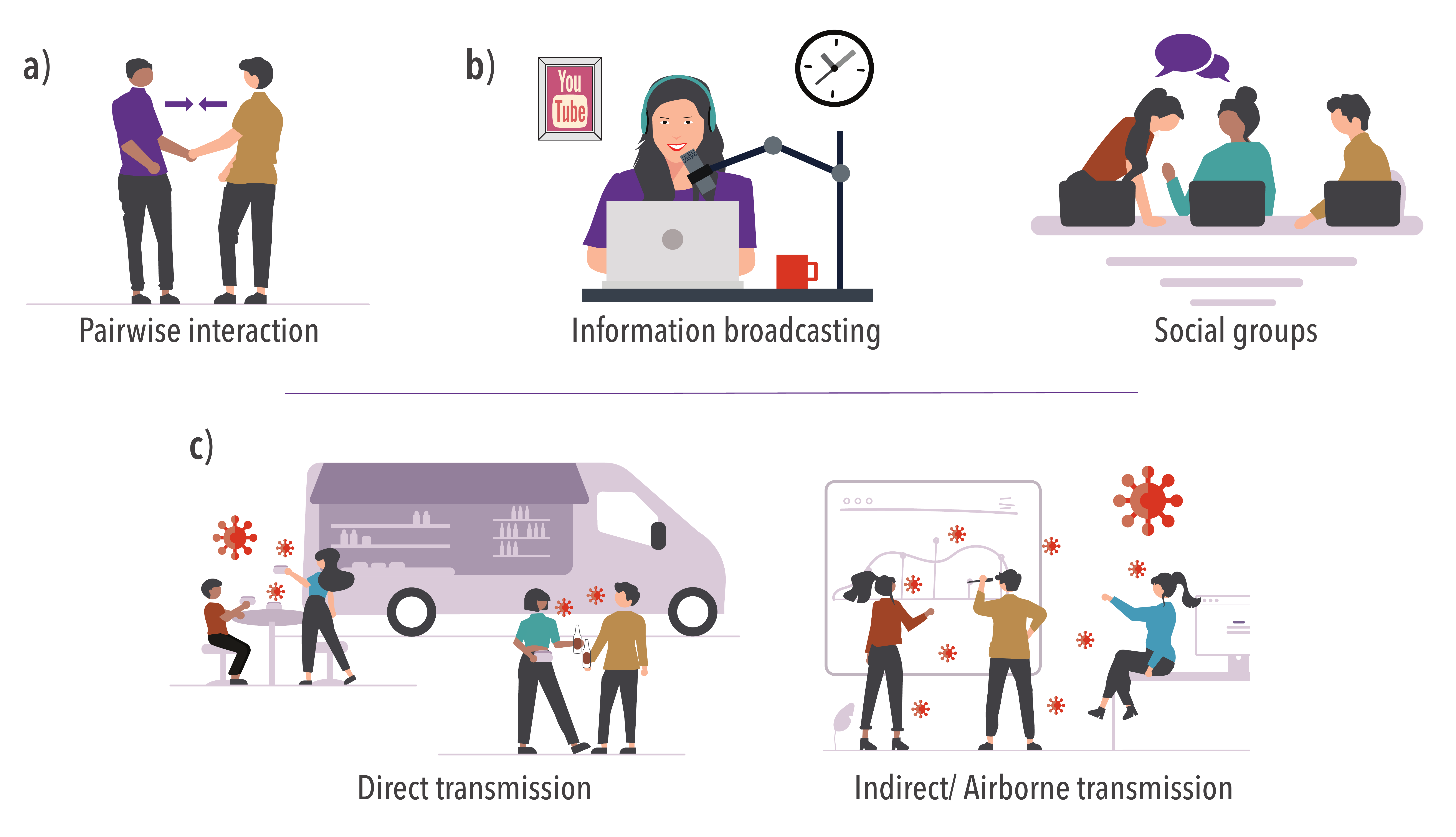}
  \caption{\textbf{Illustration of diverse interaction dynamics in social and epidemiological settings.} Traditionally, network models (a) operated under the assumption of pairwise interactions. However, this oversimplification is challenged in social scenarios (b), where exceptions arise, like the influence of public figures or intricate dynamics within cohesive social groups. Likewise, in epidemiological contexts (c), direct droplet transmission may adhere to pairwise dynamics, but in indirect or airborne transmission group interactions play a major role.}
  
\label{fig:empirical}
\end{figure*}

In this section, we introduce the phenomena motivating higher-order dynamics in the study of complex systems. Fig. \ref{fig:empirical} provides a visual representation, offering a sketch of different types of interactions, from pairwise to group interactions in different contexts.

\subsection{Sociological motivations}

The empirical investigation of coordination, norms, and behaviors spreading across networks has garnered considerable attention in the literature~\cite{Centola2015, Galantucci2005, Centola2010, Hodas2014, Aral2017, Christakis2007}. Social contagion within networks, particularly the emergence of consensus without centralized institutions, stands out as a major area of interest~\cite{Sugden1989, Bikhchandani1992, Ehrlich2005, Young2015, Everall2023, Baronchelli2018}. Understanding the dynamics of social contagion and change is intricate due to the diversity of problems that span across many domains and contexts. Yet, certain commonalities prevail.

Tipping points hold significant interest within this context. Described as a threshold at which a small quantitative change in the system can trigger a nonlinear process that leads to a different state of the system~\cite{Milkoreit2018}, these points are central to the critical mass theory. This theory posits that a minority of committed individuals, upon reaching a critical size, can overturn a social convention~\cite{Centola2015}. The concept finds ample validation in theoretical models~\cite{Szymanski2011, Mistry2015, Niu2017, Baronchelli2018} and empirical studies~\cite{Kanter1977, Drude1988, Grey2006, cencetti2021temporal, o2022anatomy, Centola2018, Amato2018, Diani1992, Iacopini2022}. Intriguingly, observed critical mass thresholds span several orders of magnitude, ranging from $25\%$ to $40\%$ in some observational studies on social conventions~\cite{Kanter1977}, to as low as $0.3\%$ for linguistic norm changes in English~\cite{Amato2018, Iacopini2022}, or even encompassing just a few individuals relative to the population size in social movements~\cite{Diani1992, Iacopini2022}.

A pertinent question arises concerning the mechanisms by which small groups evolve into committed minorities. Research spanning sociology, political science~\cite{Kanter1977, Granovetter1978, Drude1988, Grey2006, Centola2018}, physics, and mathematics~\cite{Szymanski2011, Mistry2015, Niu2017, Baronchelli2018, Iacopo2019, Jhun_2019, Arruda2020, Arruda2021, barrat2021social, Battiston2021, Alvarez-Rodriguez2021, higham_2021, Higham2022, Higham2022b, battiston_networks_2020, kim2022higherorder} has explored these group interactions, which are now gaining attention in the field of complex systems due to recent inclusion of higher-order interactions in contagion models. These models offer richer dynamical behaviors, including abrupt transitions, multistability, and intermittency. Understanding social contagion within an increasingly interconnected world is crucial, potentially guiding policy decisions. For instance, these models could facilitate accelerating societal changes to address societal challenges like responding to climate emergencies~\cite{Westley2011, David2018, Nyborg2016, Milkoreit2018, Otto2020, Lenton2020, Everall2023}.

\subsection{Epidemiological motivations}
\label{sec:epimoti}

Realistic epidemic models often categorize interactions into four main groups: households, schools, workplaces, and the broader community. These groupings not only serve as significant sources of transmission but also offer a more feasible target for public health interventions compared to individual-based strategies. Intriguingly, epidemiological studies have revealed widely varied per-contact transmission probabilities within these settings, with households demonstrating the highest rates~\cite{Bi2020Aug, Sun2021Jan, Ajelli2014Nov, Sun2021Jan, Zhao2022Dec}. Larger settings exhibit distinct interaction characteristics compared to smaller groupings, impacting disease transmission dynamics significantly~\cite{lePolaindeWaroux2018Dec,Althouse2020Nov}.

Moreover, the challenges posed by varying per-contact transmission probabilities across different settings are compounded by dose-response dynamics in infection exposure and emerging insights into the mode of transmission for respiratory pathogens like SARS-CoV-2. While traditional assumptions centered around large droplets or fomite-based transmission, surveillance data swiftly indicated airborne transmission as the dominant form for SARS-CoV-2 spread~\cite{Wang2021Aug,Tang2021Apr,Robles-Romero2022May,Kohanski2020Oct, Morawska2022Nov}, even in settings with close-range interactions~\cite{Kleynhans2023Jul}. The former can be easily modeled as a pairwise interaction, but classical models struggle to address the latter~\cite{Hu2013Aug,st-onge_universal_2021, Silk2022Oct}.

Households are particularly suitable for higher-order network models like hypergraphs, especially concerning airborne diseases. Despite their apparent simplicity, they exhibit compelling phenomena, notably showcasing significantly higher Secondary Attack Rates (SAR) compared to other contexts for SARS-CoV-2 and influenza~\cite{Thompson2021Aug,Qian2021May,Sun2021Jan,Koh2020Oct,Tsang2016Feb,Mousa2021Nov, Mistry2021Jan} that decreases with household size~\cite{Pollan2020Aug,ZacharyJ.Madewell2021Aug}.

Airborne transmission also amplifies the probability of Super Spreading Events (SSEs), which are influenced by various biological, social, and environmental factors~\cite{Althouse2020Nov,Lloyd-Smith2005Nov,Aleta2022Jun}. While nodes with large degrees can simulate such behavior, it is essential to consider additional heterogeneities since context also shapes these events~\cite{Tariq2020Dec, Bi2020Aug}. In fact, during the recent COVID-19 pandemic, it was observed that up to 70\% of cases did not transmit the virus to anyone, while those who did typically infected just 1 or 2 others~\cite{Adam2020Nov, Laxminarayan2020Nov,Cooper2019Sep, Sun2021Jan}.

Furthermore, social contagion elements also wield considerable influence on epidemic spread. Mask usage, which can reduce the transmission of respiratory diseases~\cite{Leung2020May,Chu2020Jun,Lai2012May}, varies greatly across countries due to cultural and psychological factors, as well as group dynamics~\cite{Badillo-Goicoechea2021Nov, Lu2021Jun, Scheid2020Sep, Betsch2020Sep,Bir2021Jan}. Similarly, vaccination uptake is also influenced by social dynamics ~\cite{Dhanani2022Jun, Cascini2021Oct}, and some studies even demonstrate that individuals explicitly take into account group dynamics - akin to critical mass processes - to decide when to vaccinate~\cite{Lazer2021Dec}.

\subsection{Other motivations}

The relevance of contagion models is not limited to the epidemiological or social context. Rumor models~\cite{daleykendall1964, makithompson1973} inspired and formed the theoretical basis for the gossip protocol~\cite{Demers1987}, a powerful paradigm used in the design of reliable and efficient decentralized distributed protocols~\cite{Montresor2017}. This protocol is widely used in peer-to-peer (P2P) networks~\cite{Maarten2007, Montresor2017, Ripeanu2002}, including the Gnutella P2P network~\cite{Ripeanu2002}, and cryptocurrency networks such as Bitcoin~\cite{Koshy2014, Zulfiker21019} or Ethereum~\cite{Kiffer2021} and their derivatives. Moreover, the application of hypergraph theory has demonstrated its efficacy in modeling wireless and 5G networks, as evident in existing literature~\cite{Zhang2016, Zhang2017, Sun2017, Nyasulu2021}. These instances suggest the potential for social contagion models in higher-order networks to inspire novel protocols and methodologies.

\section{Contagion on Higher-Order Systems}
\label{sec:Contagion}

There are many approaches to model contagion processes on higher-order structures. In this section, we unify the most common contagion models in hypergraphs using a single equation that can be adapted to capture different behaviors, which can be either social or epidemic-inspired. Importantly, we show that this equation can also be reduced to the pairwise case, emphasizing that the higher-order formulation is a generalization of the classical pairwise case.

We focus on the two most paradigmatic contagion models, the SIS, and the SIR, in hypergraphs. In the thermodynamic limit, the SIS model has a single absorbing state, while the SIR has infinitely many absorbing states. Here, we define a hypergraph, $\mathcal{H}$, as a set of vertices, $\mathcal{V} = \{ v_i \}$, and a set of hyperedges, $\mathcal{E} = \{ e_j \}$, where $e_j$ is a subset of $\mathcal{V}$ with arbitrary cardinality $|e_j|$. The number of vertices is $N = |\mathcal{V}|$. If $\max \left( |e_j| \right) = 2$ we recover a graph. If for each hyperedge with $|e_j| > 2$, its subsets are also contained in $\mathcal{E}$, we recover a simplicial complex. In the contagion models, nodes can be in one of three states: susceptible, infected, or recovered (when applicable). Note that it is common practice to adopt an epidemic-spreading nomenclature even for social contexts. To model these states, we associate each node $v_i$ with three Bernoulli random variables, $(X_i, Y_i, Z_i)$. Accordingly, susceptible individuals are in the state $(1, 0, 0)$, while infected and recovered individuals are in the states $(0, 1, 0)$, $(0, 0, 1)$, respectively.

The transition between states is defined as a collection of Poisson processes. We associate a healing mechanism to each infected node, modeled as a Poisson process with parameter $\delta$. Meanwhile, the propagation mechanism is associated with the hyperedges. Consequently, given the hyperedge $e_j$, the spreading is modeled by a Poisson process with parameter $\lambda \times \lambda^\ast(|e_j|)$, where $\lambda^\ast(|e_j|)$ is a function of the cardinality of the hyperedge and can be used to modulate the infection rate. The final component of the model is a function that captures interactions among individuals within a hyperedge. Specifically, the function $f_j^i (\{ Y\})$ models how the hyperedge $e_j$ affects the state of the node $v_i$. Note that the argument of such a function is the (infected) state of all nodes in the hypergraph, here denoted by the set $\{ Y\}$. Under these assumptions, the exact form of a SIS model in hypergraphs is
\begin{align}
    \label{eq:exact}
    \dfrac{d \E{Y_i}}{dt} = \E{-\delta Y_i + \lambda \sum_{j:v_i \in e_j} \lambda^\ast(|e_j|) X_i f_j^i (\{ Y\})},
\end{align}
where $\E{\cdot}$ is the expectation operator, $\lambda$ can be thought of as the control parameter, and $\lambda^\ast(|e_j|)$ is a local parameter that weights each hyperedge differently depending on its cardinality (i.e., the group size). Notably, while $\lambda^\ast(|e_j|)$ could be absorbed into $f_j^i (\{ Y\})$, it has been left out to emphasize the individual contribution of each type of hyperedge to the process. To describe the SIR model, an additional equation is needed:
\begin{equation}
	\label{eq:exactR}
	\dfrac{d \E{Z_i}}{dt} = \E{\delta Y_i}.
\end{equation}

To solve Eq.~\eqref{eq:exact} in its exact form for an arbitrary hypergraph, we need to solve an ODE system with $2^N$ equations, which is not feasible in most cases. However, in some cases, one can use structural symmetries to reduce the system size to $N$ equations. For the simplicial contagion model, such an approach was formulated in a complete hypergraph~\cite{kiss2023insights}, and for the social contagion model based on critical masses, such an approach was formulated in~\cite{Arruda2023} for a homogeneous structure.

A few papers have presented some results for contagion models using general interaction functions. The first SIS model on hypergraphs was published in~\cite{Bodo2016}, where the authors considered a general formalism, studying both the exact equations and a mean-field approach. Considering concave functions $f$, in~\cite{higham_2021}, results were obtained for the spectral thresholds for extinction, where the authors derived a spectral bound for the expected time to extinction and spectral conditions for the local and global stability of the zero-activity state. Some of these results were also extended to non-concave functions. In~\cite{Higham2022b}, temporal hypergraphs were considered, and a spectral threshold for the spreading rate below which the activity dies out was obtained in terms of a static expectation matrix, which is an expected clique expansion of the hypergraph.

\begin{figure}[t!]
  \centering \includegraphics*[width=\columnwidth]{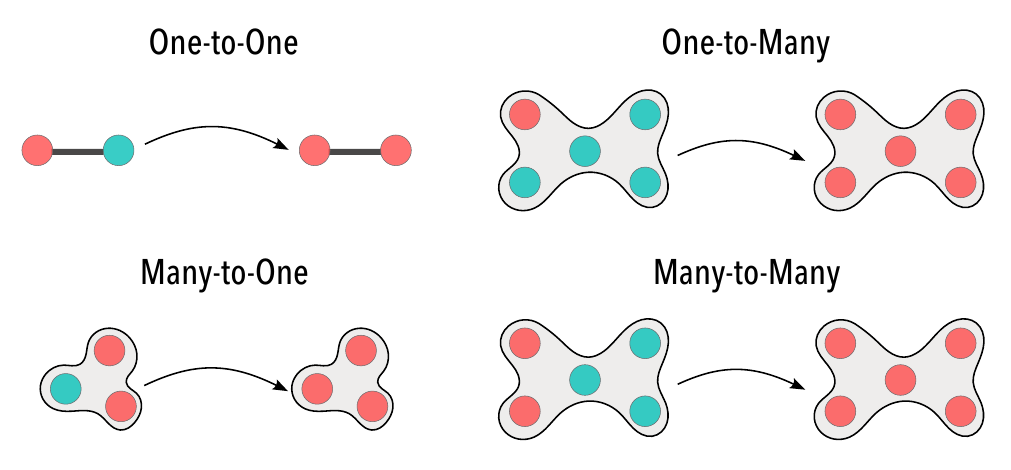}
  \caption{\textbf{Schematic representation of the different types of interactions in higher-order networks.} Functions are separated by how many individuals are required to make the hyperedge active and how many individuals are affected by that function. Here, red nodes represent infected or active individuals, while green nodes represent healthy or inactive individuals. For example, the pairwise interactions are of the form one-to-one. Note, however, that we can also have higher-order one-to-one interactions when the spreading rate depends on a nonlinear function of a node's infected neighbors. The other schemes are higher-order, and their main difference is the number of individuals needed to trigger the spread. This simple classification allows us to easily distinguish between different models.}
\label{fig:Functions}
\end{figure}

In the next sections, we revise some of the most popular spreading models in hypergraphs and simplicial complexes and show how they can be obtained from Eq.~\eqref{eq:exact}. We also present some generalizations, results, and perspectives.
Fig.~\ref{fig:Functions} is a graphical representation of the different types of interactions in higher-order networks. The one-to-one setting describes the pairwise model and the power-law contagion kernel presented in Sections~\ref{sec:sis} and~\ref{sec:pl}, respectively. The many-to-one is used in the simplicial contagion model in section~\ref{sec:simplicial}, while the many-to-many is used by the critical mass threshold model described in section~\ref{sec:sisCM}. The many-to-one interaction type can be modeled as a special case of the critical mass threshold model and has also been studied in~\cite{Landry2020}. We note that the one-to-one does not necessarily imply a lower-order interaction.

\subsection{The pairwise SIS and SIR}
\label{sec:sis}

These models have recently been reviewed in~\cite{Pastor-Satorras2015, Arruda2018} and are outside the scope of this review. However, for the sake of comparison, it is instructive to recall a few results of phase transitions in graphs, such as localization properties, which are of particular interest for higher-order systems~\cite{St-Onge2021a, St-Onge2021b}. Furthermore, the standard SIS pairwise model can be recovered from Eq.~\eqref{eq:exact} by considering
\begin{equation}
    f_j^i (\{ Y\}) = Y_j,
\end{equation}
and setting the maximum cardinality of $\mathcal{H}$ to two. With Eq.~\eqref{eq:exactR}, we would also recover the SIR pairwise model, but for the rest of the section, we will focus on the SIS.

\subsubsection{Behavior Observed}

In a homogeneous network, the transition is continuous, and the critical point is finite and non-zero. But one of the most important results of epidemic spreading in networks is the existence of a vanishing critical point for networks where the second moment of the degree distribution diverges under the mean-field approximation~\cite{Vespignani2001}. For this reason, power-law networks, $P(k) \sim k^{-\gamma}$, attracted particular interest as the second moment diverges for $2 < \gamma < 3$. More generally, depending on the network characteristics, the transition can be driven by different activation mechanisms, namely: (i) collective, (ii) $k$-core, or (iii) hub~\cite{Pastor-Satorras2012, Boguna2013, Cota2018}. From a physics perspective, in the collective case, the whole network is active after the critical point. In the second case, the process is localized in the core of the network. Lastly, in the hub activation mechanism, the process is localized around the hubs. For multilayer networks, a similar argument has been made about the localization properties of the phase transition. However, in this case, we have additional scenarios. Here, the transition can be delocalized or layer-localized~\cite{Arruda2017}. When layer-localized, these scenarios for simple graphs may also apply.

Note that discontinuous phase transitions can also be found in epidemic processes in graphs, but only in very specific cases. One example is a disease spreading in adaptive networks, where agents change their connections depending on their state, as in~\cite{Gross2006, Scarpino2016}. For a review of this topic, see~\cite{Gross_review2008}. The motivation in~\cite{Scarpino2016} is teachers; if they are infected, they can be replaced. However, this seemingly rational behavior backfires if the spreading rate is sufficient to infect the students before the teacher leaves. In this case, the movement of teachers leads to a higher proportion of infected individuals, as more and more teachers become infected and carry the disease home with them. This implies a discontinuous transition and hysteresis. Another example would be cooperative diseases~\cite{Chen2017}.

\subsubsection{Analytical Approaches}

\begin{figure}[t!]
  \centering \includegraphics*[width=\columnwidth]{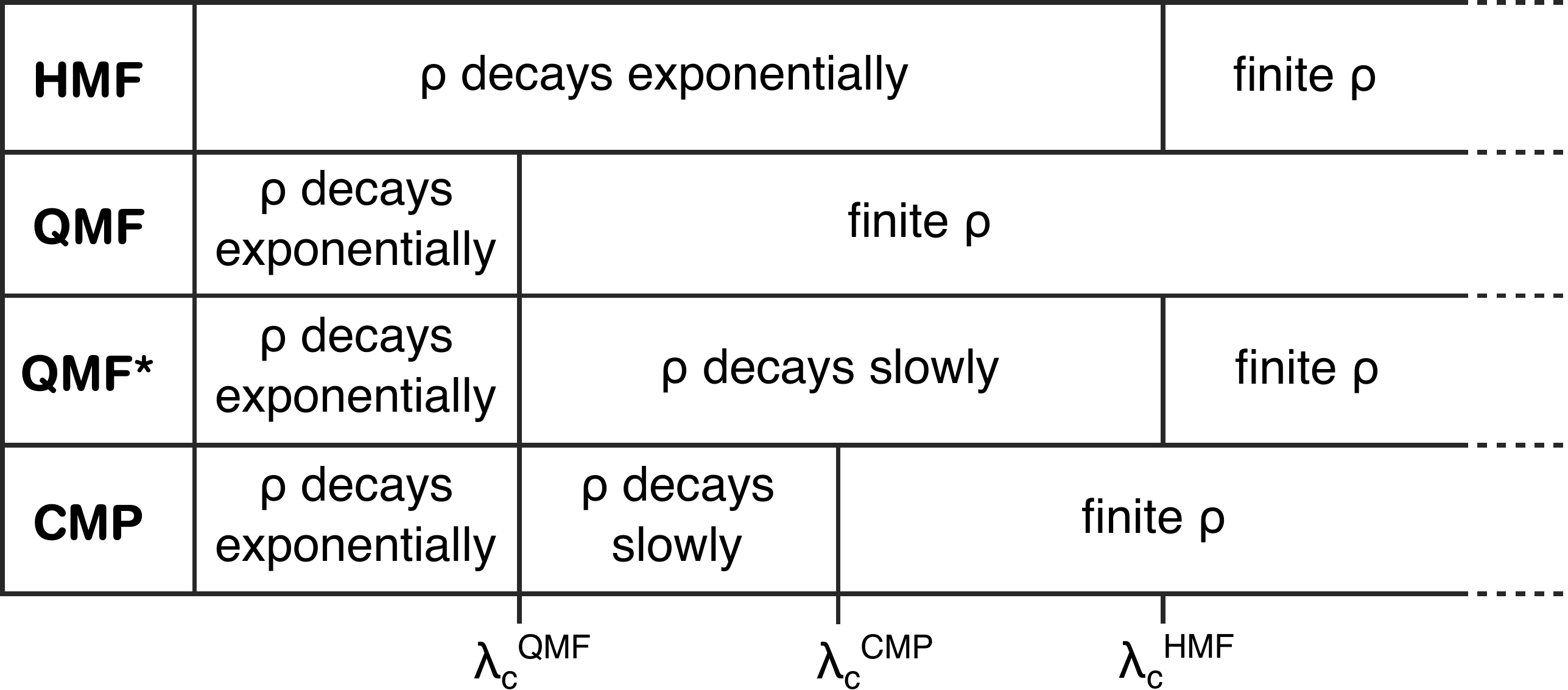}
  \caption{\textbf{SIS prevalence $\rho$ in the pairwise case according to some approaches.} We present a summary of the predictions of some of the most common approaches, the heterogeneous mean-field (HMF), the quenched mean-field (QMF), and the cumulative merging percolation (CMP). In addition, $\lambda_c^{HMF}$, $\lambda_c^{QMF}$, and $\lambda_c^{CMP}$ denote the critical point for the QMF, HMF, and CMP theories, respectively. Here, QMF* stands for the QMF theory as reinterpreted in~\cite{Goltsev2012, Lee2013}. Figure adapted from~\cite{Pastor-Satorras2020}. We emphasize that, depending on the network structure and the network structure, different analytical approaches may have the same critical point prediction. For example, for a power-law degree distribution, $P(k) \sim k^{-\gamma}$, with $2 < \gamma < 2.5$, both the QMF and HMF theories predict the same critical point.}
\label{fig:phasediagram}
\end{figure}

Most of the analytical approaches have been reviewed in~\cite{Pastor-Satorras2015, Arruda2018}. In our context, the most relevant are:
\begin{itemize}
 \item Mean-field (MF) assumes that the system is completely homogeneous \cite{Vespignani2001, barrat_2008, Pastor-Satorras2015, Arruda2018}.

 \item Heterogeneous mean-field (HMF), also called degree-based mean-field (DBMF), assumes statistical equivalence among nodes with the same degree, neglecting dynamical correlations and partially structural correlations~\cite{Vespignani2001, barrat_2008, Pastor-Satorras2015, Arruda2018}.

 \item Quenched mean-field (QMF), also called individual-based mean-field (IBMF) or N-intertwined mean-field approach (NIMFA), considers structural correlations while neglecting correlations between individual states~\cite{Mieghem2009, Mieghem2014performance,Pastor-Satorras2015, Arruda2018}.

 \item Pair-quenched mean-field (PQMF) considers both structure and dynamical, second-order correlations~\cite{Cator2012, Mata_2013}.

 \item Approximate master equations (AME), which consider the state of nodes and their immediate neighbors, generating large systems of differential equations~\cite{Dufresne2010, OSullivan2015, Marceau2010, Gleeson2011, Gleeson2013}.

 \item Discrete-time Markov chain approaches, also called microscopic Markov chains (MMC)~\cite{Gomez_2010}. In this approximation, structural correlations are considered, dynamical correlations are neglected, and time evolves in discrete steps. This can be regarded as a discrete-time version of the QMF.

 \item Epidemic Link Equation (ELE), which can be interpreted as the discrete-time version of the PQMF~\cite{Matamalas2018}.
\end{itemize}

An important result not covered in these recent reviews is the use of cumulative merging percolation (CMP) to study the critical properties of the SIS process, which provides an explanation of the mechanisms behind the phase transition~\cite{Pastor-Satorras2020}. The behavior predicted by each theory is summarized in Fig.~\ref{fig:phasediagram}.

\subsection{The SIS on hypergraphs}
\label{sec:sisH}

As previously mentioned, the first SIS model on hypergraphs was published in~\cite{Bodo2016} and was motivated by the spread of epidemics in household structures, workplaces, and schools. In this model, they assume that the infection pressure on susceptible individuals is not proportional to the number of infected individuals, so that:
\begin{equation}
\label{eq:step}
    f_j^i (\{ Y\}) =
    \begin{cases}
        m \qquad &\text{if } m < c \\
        c & \text{otherwise}
    \end{cases},
\end{equation}
where $m = \sum_{k: v_k \in e_j; v_k \neq v_i} Y_k$. In addition to the exact formulation using the Kolmogorov equations, the authors also proposed a mean-field analysis and concluded that their mean-field approximation performs well for regular random hypergraphs. Moreover, when considering a structure that includes households and workplaces, the mean-field solution grows faster than the Monte Carlo solution, but their steady-state solutions are close.

Aiming to model non-pharmaceutical interventions in realistic scenarios, the authors in~\cite{Antelmi2021} proposed a temporal hypergraph approach that extends the model proposed in~\cite{Bodo2016} by considering both direct (person-to-person, i.e., pairwise interaction) and indirect contacts (infection through an intermediary, a contaminated environment). Their approach is mainly computational, based on agent-based simulations in which different interventions are evaluated. Their results emphasize the role of personal protection and hygiene measures in slowing down the spread.

\subsection{The Simplicial Contagion Model}
\label{sec:simplicial}

Aiming to model social contagion processes such as opinion formation or adoption of novelties, where complex influence mechanisms and reinforcement are present, in~\cite{Iacopo2019} the authors proposed to use simplicial complexes, a particular type of hypergraphs, and a multiplicative interaction function. From Eq.~\eqref{eq:exact}, the simplicial contagion model can be obtained using
\begin{equation}
    \label{eq:product}
    f_j^i (\{ Y\}) = \prod_{k:v_k \in e_j; v_k \neq v_i} Y_k,
\end{equation}
and replacing the hypergraph $\mathcal{H}$ by a simplicial complex. Note that the model still holds for an arbitrary hypergraph, as discussed in~\cite{barrat2021social}.
Therefore, in this section, we summarize the results for both structures.

\subsubsection{Behavior Observed}

In~\cite{Iacopo2019}, a mean-field approach was proposed for an arbitrary order simplicial complex. The authors focused their efforts on the analysis of the simplicial complex with dimension two, i.e., triangles and pairwise edges. They were able to show analytically and numerically a discontinuous phase transition with a hysteresis loop.
Analytically, the origin of such a transition is the existence of a third-order polynomial equation, which means that we could have two stable solutions separated by an unstable one.

Regarding the phase transition from the disease-free (inactive) state to an endemic (active) state, the interaction function in Eq.~\eqref{eq:product} has also been studied in uniform hypergraphs with a power-law degree distribution in~\cite{Jhun_2019}. Using an HMF, the authors showed that continuous or hybrid transitions occur when the hub effect is dominant or weak, respectively. In addition, critical exponents were calculated analytically and validated numerically. Interestingly, in~\cite{Arruda2021}, using the QMF approach, the authors found that a continuous phase transition can necessarily only exist if there are enough pairwise interactions. This result is in agreement with~\cite{cisneros-velarde_2021}, where a QMF approach was studied analytically, formalizing the critical and tri-critical points. This result contradicts the continuous phase transitions observed in~\cite{Jhun_2019}. However, it should be noted that analytically, in~\cite{Jhun_2019}, an HMF approach was used where the network is annealed, which is different from the QMF assumptions in~\cite{Arruda2021}. Also, numerically, the simulations in~\cite{Jhun_2019} start from a fully infected population that is not sufficiently close to the disease-free state and thus does not satisfy the assumptions of the linear stability analysis used in~\cite{Arruda2021}, where near the disease-free state the probability of having $(e-1)$ infected nodes is negligible.

Beyond the transition point, higher-order structures have been shown to have less influence in the initial phase of spreading~\cite{Li2021}, where it is difficult or almost impossible for the process to gain prevalence if only higher-order interactions are present. This implies that pairwise interactions are necessary to activate higher-order structures. Then, after this period, the higher-order structures will accelerate the spreading, making it converge faster to the stable (or metastable in the case of finite structures) state~\cite{Li2021}.

Finally, multistability was first found in the context of social contagion in higher-order structures in the critical mass threshold model~\cite{Arruda2023} (see section.~\ref{sec:sisCM}). In this case, the ingredient that produced this behavior was structural heterogeneity in the form of community structure. However, this is not the only feature that generates multistability. In fact, heterogeneity in the propagation parameters in different orders of interactions may be sufficient to have multistability in a complete simplicial complex~\cite{kiss2023insights}.

\subsubsection{Analytical Approaches}

The mean-field approach proposed in~\cite{Iacopo2019} disregards any structural and dynamical correlations. Alternatively, one can use the pair-based approximation, which explicitly describes the average correlations by taking into account the product between two random variables. This approximation is more accurate than the standard mean-field model on the random simplicial complex model~\cite{malizia2023pairbased}. Moreover, the pair-based approximation predicts a slightly smaller bistable region when compared to the standard mean-field model.

The HMF approach was first applied to social contagion in hypergraphs in~\cite{Landry2020}, where the critical point for the disease-free state (inactive or absorbing) only depends on the pairwise interactions. Similar but slightly different results were also obtained in~\cite{Xijian2024}, using a dimensionality reduction technique. Interestingly, the same theory also predicts that increasing the heterogeneity in the pairwise interactions postpones the onset of bistable behavior~\cite{Landry2020}.

The QMF technique was used in the critical mass social contagion model in~\cite{Arruda2020} (see also Sec.~\ref{sec:sisCM}). Note that this approach can be derived from equations~\ref{eq:cm} (see below) by setting $\Theta_j = |e_j| -1$ to recover the simplicial contagion model and neglecting correlations in $f_j^i$, which is a QMF requirement. We leave the discussion of this approach to Sec.~\ref{sec:sisCM}.

Concerning discrete-time approaches, in higher-order systems, both the MMC and the ELE have been proposed in~\cite{Matamalas2020} and have shown better agreement than the MF approaches when compared to Monte Carlo simulations in random simplicial complexes. The main disadvantage of the ELE is its analytical difficulties~\cite{Matamalas2020}. Another discrete-time approach is the network clique cover approximation, called the microscopic epidemic clique equation (MECLE), which uses a particular edge clique cover to account for dynamical correlations~\cite{Burgio2021}. In~\cite{Burgio2021}, the authors showed that the MECLE usually performs better than the ELE and the MMC. The disadvantage of this approach is its computational complexity since it describes the system by $N + \sum_{n=2}^m (2^n - n -1) C^{(n)}$ equations, where $C^{(n)}$ is the number of projected cliques with $n$ nodes~\cite{Burgio2021}.

The AME approximation has been generalized to describe hypergraph contagion in~\cite{St-Onge2021a, St-Onge2021b, st-onge_influential_2022}. The main advantage of this formulation is its analytical tractability, which allows for closed-form implicit expressions for the critical and tricritical points. This formulation assumes an arbitrary infection rate function and allows for an arbitrary group distribution. The results obtained with AME focus on a power-law infection kernel slightly different from Eq.~\eqref{eq:product} and thus are discussed in Sec.~\ref{sec:pl}.

The methods described above are generalizations of network approaches and, with the exception of QMF and PQMF, they neglect structural correlations. However, in higher-order systems, it is expected to find nested structures~\cite{Jihye2023}. Note that a simplicial complex is, by definition, a perfectly nested structure. To explicitly account for this feature, the facet approximation (FA) has been proposed. In the FA, this correlation is accounted for by explicitly considering a local mean-field approximation on nested structures. This is the hypergraph generalization of the clique approximation~\cite{Dufresne2010}. Interestingly, by neglecting nestedness in the FA formalism, one can obtain the same set of equations as in the HMF~\cite{Landry2020}. The accuracy of the FA has been evaluated in a random model that interpolates between a completely nested hypergraph, i.e., a simplicial complex, and a random hypergraph~\cite{Jihye2023}. In a completely nested hypergraph, the FA predicted the transition points better than the ELE, HMF, and MF approaches. Moreover, the FA on fully nested hypergraphs with only pairwise and triadic interactions predicts that infectious diseases can spread with lower pairwise infectivity, i.e., an increase in hyperedge-nestedness lowers the invasion threshold by promoting triangular infections. It also predicts the bistable regime when the triadic spread rate is large enough.

\subsubsection{Model Variations}

Several modifications of this functional form have been proposed to incorporate more realistic scenarios. Here, we mention some variations and approaches. The SIR model has been proposed in a simplicial complex in~\cite{Palafox-Castillo2022}, where the authors use a homogeneous mean-field approach to describe the process and obtain an expression for the critical point. Another model studied in simplicial complexes is the SIRS process, which is a combination of the SIS and SIR models. In this scenario, in addition to the discontinuous transitions and the bistability, the SIRS model also presents a stable limit cycle~\cite{Wang2021}. Similar effects have also been observed in the presence of births and deaths, where a steady periodic outbreak emerges under certain conditions~\cite{Leng2022}. Moreover, based on the model in~\cite{Wang2021}, a fractional SIRS model on simplicial complexes has been proposed in~\cite{Zhou2022}, which accounts for time delays caused by the latent and healing periods. In this case, a Hopf bifurcation occurs when the delay is larger than a critical value. A less common model that has also been put forward in this context is the SIWS~\cite{cui2023}. In this case, a ``water'' compartment is an infection reservoir modeled as a hyperedge accounting for indirect transmission~\cite{cui2023}.

The mean-field approaches presented in previous sections aim to describe the process using a deterministic description of the mean. An alternative approach would be to use stochastic differential equations to model unpredictable or random interactions. The simplicial complex SIS has been studied under these assumptions in~\cite{Tocino2023, Serrano2023}, where the stability of the origin was characterized, and the parameter space was partitioned into unstable, bistable, and globally asymptotically stable regions~\cite{Serrano2023}.

Temporality is another key element that has been incorporated into this class of models. This feature has been incorporated into simplicial complexes using the MMC approach by considering only the neighbors and triangles that are active at a given time~\cite{Chowdhary2021}. Focusing on homogeneous random temporal hypergraphs constrained to pairs and triangles, under the assumption that there are no correlations in the temporal structures, the effect of the higher-order contagion parameter was found to be much weaker compared to the static case~\cite{Chowdhary2021}.

It is also possible to incorporate distrust dynamics in these models by adding directionality to simplicial complexes. In~\cite{dekemmeter2023}, the authors showed that when edge signs were randomly assigned and maintained during a group interaction, increasing distrust can change the nature of the transition from discontinuous to continuous by making the bistability region associated with the first-order transition vanish. On the other hand, if the distribution of signs in the triadic groups has been biased to account for social balance theory, contagion is determined by the relative proportions of balanced and unbalanced triangles and by which configuration within these two classes is more common.

Rather than adding more realism in the interactions, another line of research focuses on studying the problem of interacting processes. An example is the coevolution of information and disease spread. Such a process can be modeled using a multiplex approach, where the information spread is modeled by simplicial complexes, while the disease is modeled by pairwise or higher-order interactions~\cite{Chang2023, Fan2022, Liu2023, LiCai2023, Wang2022,Fan2022b,Sun2022,You2023,Hong2023}. Still considering interacting processes but outside the multiplex framework, in~\cite{Lucas2023}, the authors focused on how a simplicial contagion could drive a simple contagion. They showed that above a critical driving force, the simple contagion could exhibit both discontinuous transitions and bistability. They also showed that unidirectional coupling processes between a higher-order contagion and a simple contagion can impose a discontinuous transition and hysteresis in the simple contagion.

While the former works mostly focus on the interaction between information and an epidemic process, there is also a lot of interest in studying competing pathogens~\cite{Wang2019}. In~\cite{Li2022}, they considered two competing simplicial SIS epidemics and obtained a phase diagram with nine regions. An MMC describing this process has been developed in~\cite{Nie2022}, where more accurate results are expected. In addition, a relevant feature of this type of process is homophily, which can be broadly defined as the tendency to associate and bond with similar individuals~\cite{Nie2022b, Veldt2023}. In~\cite{Nie2022b}, the authors extended an MMC to study simplicial competitive spreading dynamics between two states in the context of heterogeneous populations and homophily effects. Such an  MMC approach was also used for the SIS model in~\cite{Xue2022} and the SIR in~\cite{LiNie2022}. Finally, competitive spreading was also dealt with in~\cite{gracy2023competitive}.

As a side note, it is also possible to model deactivation as a group-based process by incorporating higher-order terms in the healing mechanism. In~\cite{Landry2020}, this mechanism was called the ``hipster effect'', motivated by the fact that if a trend is popular, then individuals may be less likely to adopt such a trend. With this modification, they found that the phase diagram exhibited a small band of bistability separating the regions of no infection and a single infected state.

\subsection{The Power-Law Infection Kernel}
\label{sec:pl}

In~\cite{st-onge_universal_2021}, the authors used COVID-19 data to challenge the assumption that there is a linear relationship between the number of infectious contacts and the risk of infection. Based on the data, they propose a power-law infection kernel defined as
\begin{equation} \label{eq:plk}
    f_j^i (\{ Y\}) = \left( \sum_{k:v_k \in e_j; v_k \neq v_i} Y_k, \right)^{\nu},
\end{equation}
where $\nu$ can modulate the non-linearity of the process. In particular, when $\nu=1$, we recover the linear case (pairwise), while social reinforcement and inhibition can be modeled by $\nu > 1$ and $\nu < 1$, respectively. Furthermore, if $\nu = 1$, and $\lambda^*(|e_j|) = |e_j|^{-\eta}$, where $\eta \in [0, 1],$ we recover the model analyzed in~\cite{St-Onge2021a, St-Onge2021b}, where the same authors used a bipartite representation in their mean-field description of the model.

\subsubsection{Behavior Observed}

In terms of sub-extensive localization (also called mesoscale localization in~\cite{St-Onge2021a, St-Onge2021b, st-onge_influential_2022}), the behavior is driven by the most influential groups. Localization also affects the phase diagram, with the effects being amplified by superlinear infection ($\nu > 1$). In this case, the critical point scales as $\lambda_c \sim k_{\max}^{-\nu}$, and for $\lambda$ near $\lambda_c$, the infected nodes are concentrated in the largest groups. This localization pattern inhibits bistability by forcing an endemic state with a very small global fraction of infected nodes~\cite{st-onge_influential_2022}.

For the active state, in~\cite{st-onge_universal_2021}, the authors considered the problem of maximizing influence. Focusing on the early stages of the spreading, they proposed two strategies for allocating initial seeds to influential spreaders or to influential groups. They showed that the group-based strategy tends to perform better for sufficiently nonlinear processes.

In a similar spirit, a relationship between core decomposition and SIS and SIR-like contagion processes has been studied in~\cite{mancastroppa2023hypercores}.
Based on the concept of $(k,m)$-bipartite core decomposition~\cite{Ahmed2007, Monika2015, Liu2020}, a family of hypercore centralities was defined, and two versions were proposed: (i) the size-independent hypercore, and (ii) the frequency-based hypercore. Nodes inside cores with either higher degree, $k$, or cardinality, $m$, often tend to be more infectious during the SIS process, implying that the process is expected to be more localized in this region of the hypergraph. Interestingly, in the supplementary material of~\cite{mancastroppa2023hypercores}, the authors show that their results are also valid for the critical mass threshold model (Sec.~\ref{sec:sisCM}). Moreover, considering the naming game in higher-order structures~\cite{Iacopini2022}, nodes inside the inner cores may be particularly efficient at overturning a majority convention if they belong to a committed minority.

Crucially, we should mention that the core decomposition studied in~\cite{mancastroppa2023hypercores} is different from the percolation process with a similar name in~\cite{lee2023kqcore,bianconi2023nature}, where a hybrid phase transition was found and characterized. The main difference is that in the bipartite core decomposition, the hyperedges that may be in an inner core may not be in the original hypergraph. This does not happen in the hypergraph core decomposition~\cite{lee2023kqcore,bianconi2023nature}.

\subsubsection{Analytical Approaches}

In~\cite{st-onge_universal_2021}, the authors used an HMF. This theory predicts a discontinuous phase transition, super-exponential scattering, and hysteresis. Alternatively, a group-based AME has been proposed to study this type of process in~\cite{St-Onge2021a, St-Onge2021b, st-onge_influential_2022}. The main advantage of this approach is its analytical tractability, allowing closed expressions for the critical and tricritical points. It is worth noting that the AME approach was developed for an arbitrary spreading function. Focusing on the power-law kernel, Eq.~\eqref{eq:plk}, AME showed that a large third moment of the cardinality distribution suppresses the discontinuous phase transitions with a bistable regime~\cite{st-onge_influential_2022}.

\subsection{The Critical Mass Threshold Model}
\label{sec:sisCM}

The critical mass processes studied in the social and political sciences motivated~\cite{Arruda2020, Arruda2023} to propose the propagation function:
\begin{equation}
    \label{eq:cm}
    f_j^i (\{ Y\}) = H \left(\sum_{k:v_k \in e_j; v_k \neq v_i} Y_k -  \Theta_j \right),
\end{equation}
where $H(\cdot)$ is the Heaviside function and $\Theta_j$ is a positive integer.
Note that the simplicial contagion model in Sec.~\ref{sec:simplicial} can be obtained by setting $\Theta_j = |e_j| - 1$ and using the appropriate hypergraph. Similarly, with $\Theta_j = 1$, one would recover the so-called individual contagion, where nodes within a hypergraph are activated if at least one node is active~\cite{Landry2020}.

\subsubsection{Behavior Observed}

A hybrid phase transition has been observed in this class of models. In a random regular graph with a single hyperedge covering each vertex (hyperblob), this transition was characterized using an exact formulation and finite-size analysis~\cite{Arruda2023}. This structure is probably not representative of real systems, but it provides an argument in favor of hybrid phase transitions in this model. Furthermore, this is consistent with the susceptibility curves observed in real and artificial hypergraphs as shown in~\cite{Arruda2020, Arruda2023}. These results also agree with the results for uniform hypergraphs with a power-law degree distribution in~\cite{Jhun_2019}, where the authors showed that continuous or hybrid transitions occur when the hub effect is dominant or weak, respectively (see also Sec.~\ref{sec:simplicial}).

In spite of the critical behavior, this model also presents multistability and intermittency~\cite{Arruda2023}. Both are related to the presence of a community structure. In the former, there can be multiple states for the same set of dynamical parameters, and a different state is reached depending on the initial condition. It is worth noting that multistability was also predicted in a complete simplicial complex with a specific rate distribution~\cite{Iacopo2019}. Interestingly, according to~\cite{Young2015}, an important implication for the way norms evolve is the presence of multiple equilibria.

Regarding intermittency, the results in~\cite{Arruda2023} suggest that when bridges (hyperedges connecting two different communities) are scarce, the communities are dynamically disconnected. Then, we may have multiple stable solutions for the same value of $\lambda$. As we add bridging hyperedges, we allow the process to move across communities. However, this can destroy the multiple stable solutions by merging them into a bimodal distribution of states and creating intermittency. We note that a similar effect was also observed by increasing/decreasing the bridges' hyperedge cardinalities and by changing the critical mass threshold $\Theta^*$.

Moreover, the same results for the relationship between core decomposition mentioned in Sec.~\ref{sec:simplicial} were also valid for the critical mass threshold model. In particular, it was shown in~\cite{mancastroppa2023hypercores} that the time to reach the metastable state for the SIS-like process is shorter when the initial seeds are in the inner cores. Also, the final fraction of recovered individuals for the SIR-like process tends to be higher when the seed nodes are placed in the inner core.

\subsubsection{Analytical Approaches}

In~\cite{Arruda2020}, a QMF approach was proposed, and a closed expression for the critical point was presented for the hyperstar (a star graph with a single hyperedge covering each vertex) and the hyperblob. These results were extended to a general hypergraph in~\cite{Higham2022}, where the stability of the disease-free (inactive state) was studied, and both global and local stability conditions were derived. Finally, the QMF approach could also capture multistability~\cite{Arruda2023}. However, the intermittent behavior is not captured by this approximation. The QMF captures the peaks of the state distribution as if they were metastable states~\cite{Arruda2023}. This is indeed expected since the QMF neglects correlations and stochastic fluctuations.

The so-called composite effective degree Markov chain approach (CEDMA)~\cite{Chen2023} is an alternative framework to analyze the critical mass model in hypergraphs. The CEDMA classifies nodes according to the number of neighbors and hyperedges in different states. Numerical experiments suggest that the CEDMA presents a higher accuracy than the MMC~\cite{Chen2023}. The main disadvantage of this approach is the increasing computational cost when considering hypergraphs with higher cardinalities~\cite{Chen2023}.

\subsection{Other contagion models}

Despite the generality of the framework proposed in Sec.~\ref{sec:sisH}, there are other processes that also fit in the class of contagion models but whose formulation is not straightforward using Eq.~\eqref{eq:exact}.

In~\cite{kim2022higherorder}, the authors propose a variation of the SIR model to analyze higher-order connected components of mesoscale connected structures. The $m$-th connected component is a sub-hypergraph in which hyperedges are assumed to be connected if they share at least $m$ nodes. This model could still fit in the form of Eq.~\eqref{eq:exact}, but the analysis would be somewhat more complicated.

A multistage model has been proposed to describe the spread of information driven by the spatiotemporal evolution of a public health emergency~\cite{Su2023}. This model is a variation of the SIR model, where nodes can be susceptible or infected in areas affected or not by a public health emergency. A mean-field approach and the critical value have also been studied in~\cite{Su2023}. The contagion is given by a simplicial complex with the functional form of Eq.~\eqref{eq:product}. Moreover, a delay-differential approach has been studied in~\cite{Xijian2024}, where the interaction depends on a delayed state of the active nodes. This model follows a function of the form of Eq.~\eqref{eq:product} but uses delayed states, i.e., $Y_i(t-\tau)$. Despite the model differences, the obtained critical point agrees with the results observed in other works such as~\cite{Ghosh2023}.

Digital contact tracing on hypergraphs has also ben considered~\cite{Li2022}. In this model, hyperedges can be in one of two states: traced or untraced. If the individuals in the hyperedge carry the contact tracing application, then spreading in that hyperedge is suppressed. In this study, the authors used a link percolation process to mimic SIR propagation, which is outside the models covered by Eq.~\eqref{eq:exact}. In artificial cases, the authors verified that digital contact tracing reduces the epidemic to larger cardinality of hyperedges. On the other hand, in real hypergraphs, the impact of digital contact tracing is observed to be significant for low spreading rates~\cite{Nie2023}.

Regarding immunization strategies, several strategies have been discussed in~\cite{Jhun2021}. Among them, the author studied (i) immunization of hyperedges with high simultaneous infection probability (defined as the product of the infection probabilities of the nodes in a hyperedge), (ii) a generalized version of the edge epidemic importance (EI)-based immunization strategy (originally proposed for graphs in~\cite{Matamalas2018}), and (iii) immunization of hyperedges with high H-eigenscore in uniform hypergraphs (for its definition, we refer to~\cite{Qi2005,Lim2005,qi2017tensor}). The author showed that the herd immunity threshold is slightly smaller than that of the EI-based strategy. However, it has significantly lower computational cost~\cite{Jhun2021}. On the other hand, a voluntary vaccination scheme was studied in~\cite{Nie2023b}. In this case, the SIR scheme models the spread of the disease, and the voluntary vaccination is captured using a game theory approach.

\section{Generalities, particularities, and perspectives}
\label{sec:Generalities}

As we have seen, there is no unique generalization of the SIS model to higher-order contexts. However, the common feature of the higher-order models is the presence of a nonlinear function $f_j^i (\{ Y\})$. The functional form in models~\ref{sec:sis},~\ref{sec:sisH}, and~\ref{sec:pl} was inspired by epidemic processes, whereas the models in~\ref{sec:simplicial} and~\ref{sec:sisCM} were motivated by social contexts. But if one chooses $f_j^i (\{ Y\})$ to be linear, then one can reinterpret the hypergraph as a graph with cliques and recover a weighted version of the classical SIS on networks. This argument is part of the debate about what is a higher-order system and has been discussed in~\cite{Bick2023, neuhauser2023learning, Rosas2022}.

\subsection{Theoretical perspectives}

Social contagion models on higher-order structures exhibit a wider range of behaviors compared to pairwise graphs. For example, we have described discontinuous transitions, bistability or multistability, and intermittency. However, new phenomenology may yet be found. Higher interaction orders create a unique complexity horizon, the implications of which are still largely unknown. Similarly, a systematic characterization of the phase diagram for the above functions still needs to be improved. While some efforts have started to tackle this matter, outcomes remain predominantly confined to homogeneous structures. Studies exploring heterogeneous structures, particularly in general cases, are yet to be extensively explored.

A fundamental question would be to understand how general each behavior is. Here, we have highlighted the similarities, as many models are dynamically described by Eq.~\eqref{eq:exact}, and the particularities, as each model has a different interacting function $f_j^i$. Following a linear stability analysis suggests that, at least locally, they may share many properties. An exception should be made for the function in eq.~\eqref{eq:cm}, which is not differentiable in the whole domain. Consider, for example, multistability. It has been found in the critical mass model~\cite{Arruda2023} and in the complete simplicial complex~\cite{kiss2023insights}. However, the generality of these results is unknown. In other words, is multistability present in any of the higher-order models?

Each newly identified behavior also raises the question of its generating mechanisms. Still considering the multistability, in the critical mass model, it was associated with the presence of a community structure~\cite{Arruda2023}. At the same time, it was associated with parameter heterogeneity in the full simplex case~\cite{kiss2023insights}. These are two different mechanisms that produce the same behavior. So the natural question would be, what would be the other mechanisms? Moreover, this structure has been shown to produce multiple transitions and even intermittent behavior in the critical mass model. Would the same be true for the full simplex case? And under what circumstances?

These observations lead to a more fundamental question: what are the sufficient and necessary conditions for these behaviors? As briefly discussed in section ~\ref{sec:sis}, we have a reasonable understanding of the mechanisms behind SIS and SIR behavior on graphs, something that cannot be said for the higher-order cases. However, given that the graph is only a special case of the hypergraph, a general theory should naturally extend these results. There are a number of analyses moving in this direction. One of them is studying the impact of localization in spreading processes, widely discussed in graphs~\cite{Arruda2017}, and also discussed in higher-order cases in~\cite{St-Onge2021b,st-onge_influential_2022}. However, a full understanding of localization, including its connection to graph cases, its generality, and the transition mechanisms it might describe, is still lacking.

We emphasize that the functions proposed in the literature probably do not cover specific cases. Thus, studying and proposing new functions is also a perspective. Yet, the functions that model the interactions may turn out to be highly context-dependent, making model validation much more difficult. The same is true for multistage models. Since this field of study is relatively new, most of the literature efforts focus on the simplest models, i.e., SIS and SIR, but as the field progresses, more realistic and specific models should be proposed. Other direct features that can be incorporated are data-rich structures via edge-dependent vertex-weighted hypergraphs~\cite{Chitra2019} and different temporal patterns, for example~\cite{Ceria2023, cencetti2021temporal}.

Notably, most of the mathematical approaches presented here are based on extensions of graph-based mean-field techniques. These methods have proven to be very useful and have advanced our knowledge of higher-order systems. Despite this progress, higher-order systems seem to exhibit different types of correlations~\cite{Veldt2023}, which may affect their accuracy. We note that a systematic analysis of the accuracy of each of the mean-field approaches presented here is yet to be done. Another perspective is that new higher-order specific techniques can be developed either to correct the mean-field approaches or to provide new ideas. An example of this would be the facet approximation proposed in~\cite{Jihye2023} (see also Sec.~\ref{sec:simplicial}).

\subsection{Data-oriented perspectives}

Models based on mechanistic principles allow us to generate testable hypotheses that can help us understand an observed phenomenon. However, most studies published so far have used observational data only as a motivation for their modeling choices, without testing any predictions on real data.

In the context of epidemic spreading, the recent COVID-19 pandemic presents a unique opportunity for model validation as extensive surveillance data has been collected globally with unprecedented precision despite its many limitations. Yet, very few works have used this data for model construction~\cite{st-onge_universal_2021, St-Onge2021a, St-Onge2021b}, and examples of model validation are scarce~\cite{Chen2024}. We believe that the questions discussed in Sec.~\ref{sec:epimoti} provide a good starting point for such endeavors such as the nonlinear relationship between population density, exposure, and infection risk~\cite{Kleynhans2023Jul, Sun2021Jan, st-onge_universal_2021, Tsang2016Feb}, the fact that larger settings exhibit distinct interaction characteristics compared to smaller groupings~\cite{Thompson2021Aug, Qian2021May, lePolaindeWaroux2018Dec, Hu2013Aug} and how group-mediated social contagion impacts mask wearing and vaccine uptake~\cite{Lu2021Jun, Badillo-Goicoechea2021Nov, Lazer2021Dec}.

Social contagion has been classically studied in laboratory experiments with important constraints imposed by the desired observable. In higher-order contexts, the challenges are even greater as, for instance, observing a discontinuous jump requires a large number of participants so that the transition can be unambiguously labeled as discontinuous. However, there are some features that may be easier to observe, such as localization or intermittency. Thus, there are still many questions to be addressed on how to design and implement such experiments.

In a broader perspective, even though there are few ready-to-use real hypergraphs, there are many relational datasets out there that may implicitly include higher-order interactions. Think of co-authorship graphs in which two authors are linked together if they collaborated on a paper. It is straightforward to generalize the data to a hypergraph if information on who collaborated on each paper is available. However, in other contexts, this may not be that easy. For instance, a graph containing social relationships may not include the context in which they were formed or where they usually meet. Along these lines, there are proposals to reinterpret existing datasets using Bayesian methods to infer higher-order data from already collected data~\cite{Young2021} or missing higher-order interactions~\cite{Contisciani2022}. But to properly validate these approaches, it would be important to take into account higher-order structures in the data collection process.

\section{Conclusion}

We have provided a selective literature review on contagion in higher-order structures, focusing on mathematical and physical language. To this end, we have provided a unified formalization of the process covering most of the models in the literature, highlighting their similarities and differences. These models have been proposed to help us understand various real-world processes, from the spread of epidemics to social contagion. However, even though the path for further theoretical research is well established, the validation of these models with data or experiments is still scarce and the path is yet unclear. Among the many challenges of the latter, we should also emphasize the interdisciplinary nature of this line of research. Although our contribution lies in mathematics and physics, as we have seen, many of the motivations behind these models lie outside these fields. Thus, extensive collaboration efforts will be required to firmly establish the use of higher-order interactions in contagion processes.

\section*{Acknowledgments}

A.A. acknowledges support through the grant RYC2021-033226-I funded by MCIN/AEI/10.13039/ 501100011033 and the European Union ‘NextGenerationEU/PRTR’. Y.M was partially supported by the Government of Arag\'on, Spain and ``ERDF A way of making Europe'' through grant E36‐23R (FENOL), and by Ministerio de Ciencia e Innovaci\'on, Agencia Espa\~nola de Investigaci\'on (MCIN/AEI/ 10.13039/501100011033) Grant No. PID2020‐115800GB‐I00. 


\appendix
\section{Structure of Networked Systems}

\begin{figure*}[t!]
    \centering
    \includegraphics[width=\textwidth]{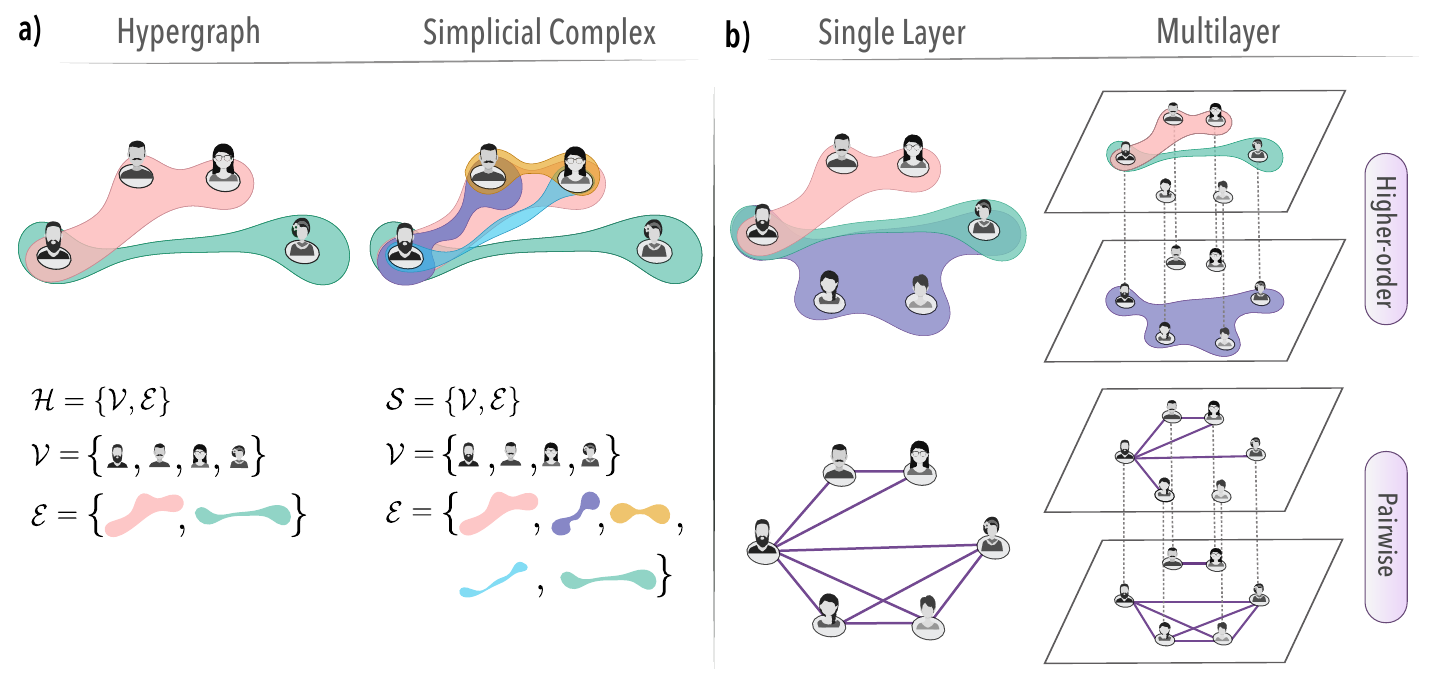}
    \caption{\textbf{Schematic representation of different networked systems.}
In (a), we compare different higher-order systems, the hypergraph, and the abstract simplicial complex. Hypergraphs consist of a set of vertices and a set of hyperedges, which are subsets of vertices. A hyperedge can contain any number of nodes. On the other hand, simplicial complexes are hypergraphs whose hyperedges respect mutual inclusion. In other words, all possible subsets of a given hyperedge must also be part of the simplicial complex. In panel (b), we exemplify different types of systems according to the type of interaction, i.e., higher-order and pairwise, and their multilevel nature, i.e., single- and multilayer settings. We note that multilayer systems are particularly suitable for modeling interacting processes, such as epidemics and information spreading. Note that this classification covers and extends previous literature.}
\label{fig:structure_examples}
\end{figure*}

A system of interacting individuals can be represented by different mathematical objects according to its type of interaction, i.e., pairwise vs. higher-order (see Fig.~\ref{fig:structure_examples} (a) for a comparison between the two main higher-order objects), or according to its multi-level organization, i.e., single-layer vs. multi-layer. For a graphical example, see Fig.~\ref{fig:structure_examples} (b) for this comparison. These objects are described formally below:

\begin{itemize}
 \item \textbf{Graph or network}: A simple graph is defined as a set of vertices connected by edges that are pairs of vertices (e.g., a friendship network).
 A non-exhaustive list of reviews and books is~\cite{Boccaletti2006, barrat_2008, Newman2010}.

 \item \textbf{Multilayer Network}: A multilayer network is a graph made up of multiple layers, each representing a different context (e.g., a friendship multilayer where we separate friends in different social circles or types of interactions (e.g., online and offline)). Multilayer networks have many subtypes. We refer to Table I in~\cite{Mikko2014} for a comprehensive classification.
 A non-exhaustive list of reviews and books is~\cite{Mikko2014, Boccaletti2014, Bianconi2018, Aleta2019}.

 \item \textbf{Hypergraph}: A graph in which hyperedges (generalized edges) can connect a subset of nodes instead of two nodes (e.g., a collection of WhatsApp groups). Formally, the hypergraph $\mathcal{H}$ is defined as a set of vertices $\mathcal{V} = \{ v_i \}$ and a set of hyperedges $\mathcal{E} = \{ e_j \}$, where $e_j$ is a subset of $\mathcal{V}$ with arbitrary cardinality $|e_j|$. The number of nodes is defined as $N = |\mathcal{V}|$, and the number of hyperedges as $M = |\mathcal{E}|$. Note that it is possible to extend the multilayer concept to hypergraphs.

 \item \textbf{Simplicial Complex}: A simplicial complex is a type of hypergraph whose set of hyperedges is complete, i.e., all possible subsets of a hyperedge are also present. In this text, when we use the term simplicial complex, we are referring to the abstract simplicial complex. Note that there is a distinction between the abstract simplicial complex, which is a hypergraph with downward inclusion, and the geometric simplicial complex, where objects are continuous. For a full discussion, we refer to~\cite{Estrada2023}. Another term for an abstract simplicial complex is an ``independence system'' used in combinatorial mathematics.
\end{itemize}

We also note that higher-order systems have been recently reviewed in~\cite{battiston_networks_2020, Torres2021, bianconi_higher-order_2021, Bick2023, Boccaletti2023}. Also, some recent perspective papers focusing on the study of higher-order systems are~\cite{Lambiotte2019, Battiston2021, Rosas2022}, and an editorial was published in~\cite{Gao2023}.

\section{Dynamical Behavior}
\label{sec:behavior}

\begin{figure*}[t!]
    \centering
    \includegraphics[width=0.9\textwidth]{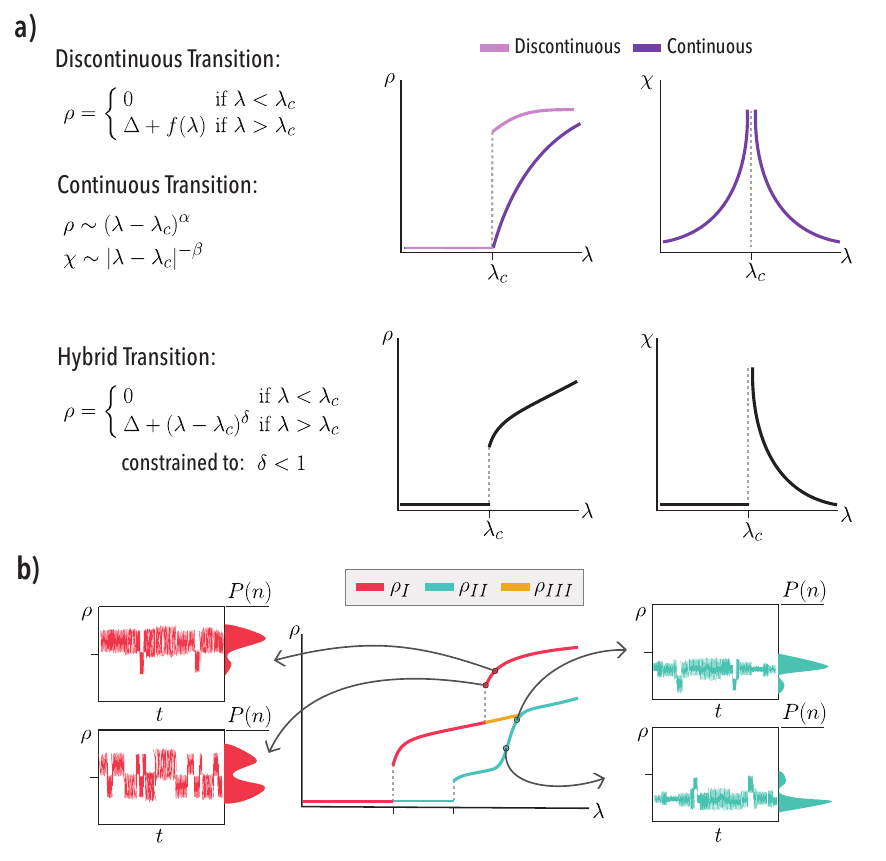}
    \caption{\textbf{Schematic representation of the most common behaviors in social contagion models.} In (a), we show a schematic representation of different types of phase transitions characterized by the order parameter and the susceptibility. The order parameter is discontinuous in a first-order phase transition, while in a second-order phase transition, the change is continuous, and the susceptibility diverges at the critical point. We can also have hybrid phase transitions where we have both the discontinuity and the divergence. In (b), we show a schematic representation of multistability, bimodal state distributions, and discontinuity. Multistability is shown in the middle panel, where for a given parameter $\lambda$, we can reach different stable solutions depending on the initial condition. The bimodal state distributions are shown in the side panels. We note that spreading processes in networks often have an unimodal bell-shaped distribution, and bimodal distributions have been observed in the critical mass model in hypergraphs. In this case, they were a consequence of the temporal intermittency (alternating periods of high and low activity) shown in the temporal evolution next to the distributions in the side panels.}
    \label{fig:phase_transitions}
\end{figure*}

Following the nomenclature of statistical mechanics, we characterize a dynamical process using two quantities:

\begin{itemize}
 \item \textbf{Order parameter}: The order parameter in the context of contagion is defined as the first moment of the distribution of the fraction of active or infected individuals, $P(n_a)$, or the average. Formally, $\rho = \E{n_a}$;
 \item \textbf{Susceptibility}: The susceptibility measures the variance of $P(n_a)$ and can be interpreted as the derivative of the order parameter. Formally, $\chi = \frac{\E{n_a^2} - \E{n_a}^2}{\E{n_a}}$, following the suggestion in~\cite{Ferreira2012}.
\end{itemize}

When a system changes from one macrostate to another, we say that it undergoes a phase transition. Note that, strictly speaking, a phase transition is defined only in the thermodynamic limit. However, we often use the analog of this concept in finite systems. This transition can occur in many different ways (for a graphical example, see Fig.~\ref{fig:phase_transitions} (a)). The ones we see more often in contagion models are:
\begin{itemize}
 \item \textbf{1st order}: Both the order parameter and the susceptibility are discontinuous at the transition;
 \item \textbf{2nd order}: The order parameter is continuous, and the susceptibility diverges at the critical point;
 \item \textbf{Hybrid Phase Transitions}: The order parameter is both discontinuous and shows scaling. Thus, the susceptibility has a one-sided divergence~\cite{Kahng2016}.
\end{itemize}

Besides phase transitions, other phenomena are relevant to contagion processes (for a graphical example, see Fig.~\ref{fig:phase_transitions} (b)):
\begin{itemize}
 \item \textbf{Hysteresis}: Hysteresis refers to the dependence of a system's state on its past. That is, for the same value of the control parameter, the state of the system will be different depending on the path followed to reach it.

 \item \textbf{Localization}: At the critical point, the transition can affect all nodes in the same way or it can be limited to some groups of nodes. Although this concept is defined at the critical point, it is often used in the supercritical regime. In such a case, the same ideas apply but it has to be measured differently.

 \item \textbf{Multistability}: More than two macrostate solutions are possible for a given set of parameters, depending on the initial condition.

 \item \textbf{Intermittency}: The system presents high and low levels of macro-activity and alternates between them. This implies a bimodal distribution of states $P(n_a)$~\cite{Arruda2023}.

\end{itemize}



\end{document}